\documentclass[conference]{IEEEtran}
\IEEEoverridecommandlockouts
\usepackage{cite}
\usepackage{amsmath,amssymb,amsfonts}
\usepackage{tikz}
 \usepackage{pgfplots}
 \pgfplotsset{compat=newest}
  \usetikzlibrary{plotmarks}
  \usetikzlibrary{arrows.meta}
  \usepgfplotslibrary{patchplots}
\usetikzlibrary{shapes,arrows}
\usetikzlibrary{decorations.pathmorphing}
\usetikzlibrary{matrix,positioning}
\usetikzlibrary{backgrounds}
\usepackage{graphicx}
\usepackage{pgf}
\usepackage{textcomp}
\usepackage{xcolor}
\usepackage[normalem]{ulem} 
\def\BibTeX{{\rm B\kern-.05em{\sc i\kern-.025em b}\kern-.08em
		T\kern-.1667em\lower.7ex\hbox{E}\kern-.125emX}}
\usepackage[letterpaper,left=0.578in, right=0.578in, top=0.7in, bottom=0.978in]{geometry}
\usepackage{algorithm}
\usepackage{algpseudocode}
\algnewcommand\algorithmicinput{\textbf{Input:}}
\algnewcommand\INPUT{\item[\algorithmicinput]}

\usepackage{bm}
\usepackage{bbm}
\newcommand{\Nt}{N_{\rm tx}}
\newcommand{\Nr}{N_{\rm rx}}
\newcommand{\Df}{\Delta f}
\newcommand{\Nrft}{N_{\rm tx}^{\rm rf}}
\newcommand{\Nrfr}{N_{\rm rx}^{\rm rf}}
\newcommand{\Tzero}{T_{\rm 0}}
\newcommand{\rect}{{\sf rect}}
\usepackage[nolist]{acronym}
\usepackage{siunitx}
\DeclareSIUnit{\belmilliwatt}{Bm}
\DeclareSIUnit{\belsquaremeter}{Bsm}
\usepackage{mathtools}
\DeclarePairedDelimiter\abs{\lvert}{\rvert}%
\usepackage{ulem}
\usepackage{xcolor}

\long\def\comment#1{}

\newfont{\bbb}{msbm10 scaled 700}

\newfont{\bb}{msbm10 scaled 1100}
\newcommand{\CC}{\mbox{\bb C}}

\newcommand{\EE}{\mbox{\bb E}}

\newcommand{\av}{{\bf a}}
\newcommand{\bv}{{\bf b}}

\newcommand{\fv}{{\bf f}}

\newcommand{\rv}{{\bf r}}
\newcommand{\sv}{{\bf s}}

\newcommand{\wv}{{\bf w}}

\newcommand{\yv}{{\bf y}}

\newcommand{\Dm}{{\bf D}}

\newcommand{\Hm}{{\bf H}}
\newcommand{\Id}{{\bf I}}

\newcommand{\Rm}{{\bf R}}

\newcommand{\Um}{{\bf U}}

\newcommand{\Ec}{{\cal E}}

\newcommand{\Psim}{\hbox{\boldmath$\Psi$}}

\newcommand{\diag}{{\hbox{diag}}}

\renewcommand{\arg}{{\hbox{arg}}}

\newcommand{\eqdef}{\stackrel{\Delta}{=}}

\newcommand{\Na}{N_{\rm a}}
\newcommand{\Nrf}{N_{\rm rf}}
\newcommand{\Ntx}{N_{\rm tx}}

\renewcommand{\H}{{\scriptscriptstyle\mathsf{H}}}

\begin{document}
	
	\begin{acronym}
		\acro{ISAC}{Integrated Sensing and Communication}
		\acro{MIMO}{Multiple-input multiple-output}
		\acro{SNR}{signal-to-noise ratio}
		\acro{mmWave}{millimeter wave}
		\acro{OFDM}{orthogonal frequency division multiplexing}
		\acro{LoS}{line-of-sight}
		\acro{ULA}{uniform linear array}
		\acro{RF}{radio frequency}
		\acro{AoA}{angle of arrival}
		\acro{AoD}{angle of departure}
		\acro{HDA}{hybrid digital analog}
		\acro{BS}{Base Station}
		\acro{UE}{User Equipment}
		\acro{KF}{Kalman Filter}
	
	\end{acronym}

	\title{Simultaneous Communication and Tracking in Arbitrary Trajectories via Beam-Space Processing}
	
	\author{\IEEEauthorblockN{
			Fernando Pedraza\IEEEauthorrefmark{1},
			Saeid K. Dehkordi\IEEEauthorrefmark{1}, 
			Mari Kobayashi\IEEEauthorrefmark{2}, 
			Giuseppe Caire\IEEEauthorrefmark{1}}
		\IEEEauthorblockA{\IEEEauthorrefmark{1}Technical University of Berlin, Germany\\
			\IEEEauthorrefmark{2}Technical University of Munich, Germany\\
			Emails: \{f.pedrazanieto, s.khalilidehkordi, caire\}@tu-berlin.de, mari.kobayashi@tum.de}}

	\maketitle

	\begin{abstract}
		
        In this paper, we develop a beam tracking scheme for an \ac{OFDM} \ac{ISAC} system with a \ac{HDA} architecture operating in the \ac{mmWave} band. Our tracking method consists of an estimation step inspired by radar signal processing techniques, and a prediction step based on simple kinematic equations. The hybrid architecture exploits the predicted state information to focus only on the directions of interest, trading off beamforming gain, hardware complexity and multistream processing capabilities.
        Our extensive simulations in arbitrary trajectories show that the proposed method can outperform state of the art beam tracking methods in terms of prediction accuracy and consequently achievable communication rate, and is fully capable of dealing with highly non-linear dynamic motion patterns.
	\end{abstract}
	
	
	\section{Introduction}
		
		\ac{ISAC} has emerged as a key enabler for 5G and beyond wireless systems to deal with the challenging requirement in terms of spectral efficiency, localization, as well as power consumption \cite{bourdoux20206g}. In particular, the waveform design for \ac{ISAC},  sharing common hardware and spectrum, has been extensively studied in the literature (see e.g. \cite{ma2019joint} and references therein). In this context, \ac{OFDM} has been widely investigated as an \ac{ISAC} waveform, in virtue of its  availability in wireless communication systems and the capability to achieve accurate radar estimation performance \cite{sturm2011waveform, SP-ISAC-Heath-Liu}. 
		
		Most of the envisioned \ac{ISAC} applications are expected to operate in \ac{mmWave} \cite{SP-ISAC-Heath-Liu}. A key feature of the \ac{mmWave} channel that motivates this design is its sparsity in the beamspace domain \cite{mmWaveChannel-Hanzo}, which connects the scattering conditions with the geometry of the environment. Due to this property, accurate channel state information is needed in order to efficiently operate in the \ac{mmWave} channel. Since acquiring this information leads to overhead, methods for dynamically inferring the channel state given past observations are currently an active area of research, often under the name of beam tracking  \cite{BayesianPredictiveBeamforming, LarewBT, VaBT, LimBT, DehkordiTracking, LiuLetter3}. Another key issue in \ac{mmWave} are the large bandwidths involved, resulting in the need for very high sampling rates and a subsequent raise in receiver complexity and energy consumption. This problem is exacerbated when large antenna arrays are considered and the received signal is sampled and demodulated at each antenna output.
		
		Most of the previous approaches to beam tracking \cite{LarewBT, VaBT, LimBT, DehkordiTracking} consider channel sounding strategies, where pilot signals are necessary to track the channel, therefore experiencing overhead. \ac{ISAC} systems eliminate this problem by considering a radar receiver co-located with the transmitter, such that the transmitted waveform can carry information while being perfectly known upon reception. However, many of the current works in this framework suffer from some practical limitations. These include (i) the use of fully digital receiver architectures incurring high energy consumption \cite{BayesianPredictiveBeamforming, LiuCaire, LiuLetter3}, (ii) the restriction to very simple trajectories such as straight roads \cite{BayesianPredictiveBeamforming, Hyun-EKF, liu2021learningbased}, and (iii) the assumption of perfect matched filtering, thus neglecting the effect of discrete sampling in time and frequency \cite{BayesianPredictiveBeamforming, LiuLetter3}.
		
		In this paper, we extend our previous work on beam refinement with \ac{HDA} architectures \cite{BeamRefinement} to a tracking scenario by combining it with recent ideas from the literature. Our method outperforms other existing approaches by showing that well designed low dimensional observations are sufficient to achieve almost optimal tracking performance, therefore alleviating the hardware complexities in \ac{mmWave}. Furthermore, we evaluate the proposed method in highly non-linear environments without the commonly used assumption of perfect matched filtering. Our simulation results demonstrate that it yields excellent performance when accurate mobility models are not available.

	\section{System Model}\label{sec:sys_model}
		
		We consider an \ac{ISAC} scenario where a \ac{BS} with $\Nt$ antennas and $\Nrft$ RF chains, co-located with a radar receiver equipped with $\Nr$ antennas and $\Nrfr$ RF chains, tries to communicate with $K$ users, while simultaneously tracking their positions over time. We let the system operate in the \ac{mmWave} band. Our model assumes that the \ac{BS} has coarse knowledge of the angular locations of the $K$ users at each measurement epoch. Such knowledge can be acquired by a beam alignment method (see e.g. \cite{Xiaoshen-singlecarrier}) if a new user enters the coverage region of the \ac{BS}, or predicted from previous observations for tracked users, as explained in Section \ref{sec:tracking}. This section focuses on a single measurement step, therefore we drop the measurement index to simplify notation.

		\subsection{Channel model}
		Assuming the $K$ users are well separated in space, we adopt the widely used mmWave radar channel model (see e.g.  \cite{SP-ISAC-Heath-Liu}) where where the received echoes arrive from the \ac{LoS} direction corresponding to each user, i.e.
		\begin{align}\label{eq:channel}
			\Hm (t, \tau) =  \sum_{k=1}^{K}h_{k} \bv(\phi_{k}) \av^\H(\phi_{k})\delta(\tau-\tau_{k})  e^{j2\pi \nu_{k} t}\,,
		\end{align}
		where $h_{k}$, $\tau_{k}$, $\nu_{k}$ and $\phi_{k}$ are respectively the complex channel coefficient, delay, Doppler and angle of arrival (AoA) of the $k$-th user. Notice that, due to the co-location of \ac{BS} and radar receiver, the angles of departure and arrival coincide. For simplicity, we focus on uniform linear arrays (ULA), such that their array response vectors have elements given by
		\begin{align}
			&[\av(\phi_{k})]_{i} = e^{j\pi(i-1)\sin(\phi_{k})},\quad i\in0,\dots,\Nt-1\\
			&[\bv(\phi_{k})]_{i} = e^{j\pi(i-1)\sin(\phi_{k})},\quad i\in0,\dots,\Nr-1
		\end{align}
		
		According to the radar equation \cite{richards2005fundamentals}, the channel coefficient $h_{k}$ satisfies
		\begin{align}
			\abs{h_{k}}^{2} = \frac{\lambda^{2}\sigma_{{\rm rcs}, k}}{(4\pi)^{3}d^{4}_{k}},
		\end{align}
		where $\lambda$ is the wavelength at the central (carrier) frequency, and $\sigma_{{\rm rcs}, k}$ and $d_{k}$ are respectively the radar cross section (RCS) and range of the $k$-th user at a given epoch.
		
		\subsection{Signal model}
		We consider \ac{OFDM} as our modulation scheme since it is one of the standardized waveforms for mmWave systems and due to its good applicability as an ISAC waveform \cite{SP-ISAC-Heath-Liu}. In particular, we use the OFDM pulse shape
		\begin{align}
			p_{n, m}(t) = \rect\left(\frac{t - n\Tzero}{\Tzero}\right)e^{j 2\pi m \Df (t-T_{\rm cp}-n\Tzero)}\,,
		\end{align}
		where $\rect(x)$ is a function taking value $1$ when $0 \leq x \leq 1$ and $0$ elsewhere, $\Df$ is the subcarrier spacing, $T_{\rm cp}$ is the cyclic prefix duration, and $\Tzero \eqdef 1/\Df + T_{\rm cp}$ is the total symbol duration including cyclic prefix. 
		Considering $N$ symbols and $M$ subcarriers, the transmitted \ac{OFDM} frame is given by
		\begin{align}\label{eq:OFDM_tx}
			\sv(t) = \frac{1}{\sqrt{K}}\sum_{k=1}^{K}\fv(\hat{\phi}_{k})\sum_{n=0}^{N-1}\sum_{m=0}^{M-1}\zeta_{k}[n, m]p_{n, m}(t)
		\end{align}
		where $\{\hat{\phi}_{k}\}$ is a set of predicted AoAs, $\fv:[-\pi/2, \pi,2]\mapsto \CC^{\Nt}$ is a beamforming function that generates a unit norm beam pointing towards the intended user, and $\zeta_{k}[n, m]$ is the $n$-th information symbol intended to user $k$ sent over subcarrier $m$. For simplicity, we considered $K \leq \Nrft$ in \eqref{eq:OFDM_tx}. 
		 As it is common in the literature \cite{OFDMvsOTFS}, we will make the assumption 
		\begin{align}
			\Df \gg \nu_{\rm max},
		\end{align}
		where $\nu_{\rm max}$ is the maximum Doppler frequency to be expected in the channel.
		
		The received backscattered signal in the antenna plane and in the absence of noise is given by 
		\begin{align}\label{eq:rx_sig_time_domain}
			\yv(t) &= \frac{1}{\sqrt{K}}\sum_{k=1}^{K}h_{k}\bv(\phi_{k})\av^\H(\phi_{k})\sv(t-\tau_{k})e^{j2\pi\nu_{k}t} \\
			&= \frac{1}{\sqrt{K}}\sum_{k=1}^{K}h_{k}\bv(\phi_{k})\sum_{k'=1}^{K}\av^\H(\phi_{k})\fv(\hat{\phi}_{k'})\nonumber\\
			&\qquad\qquad\sum_{n=0}^{N-1}\sum_{m=0}^{M-1}  \zeta_{k'}[n, m]p_{n, m}(t-\tau_{k})e^{j2\pi\nu_{k}t}\\
			&\approx\frac{1}{\sqrt{K}}\sum_{k=1}^{K}h_{k}\bv(\phi_{k})\av^\H(\phi_{k})\fv(\hat{\phi}_{k})\nonumber\\
			&\qquad\qquad\sum_{n=0}^{N-1}\sum_{m=0}^{M-1}  \zeta_{k}[n, m]p_{n,m}(t-\tau_{k})e^{j2\pi\nu_{k}t},
		\end{align}
		where the last step follows from the approximation  $\abs{\av^\H(\phi_{k})\fv(\hat{\phi}_{k'})} \approx 0$ for $k' \neq k$, which is accurate in massive MIMO systems when the users are spatially separated and the predictions are close to the true value \cite{mMIMO-fundamentals}.
		
		Aiming to reduce hardware complexity and energy consumption at the radar receiver, we process the received signal $\yv(t)$ by a reduction matrix before sampling. In order to be able to produce super-resolution angle estimates, we let the radar receiver focus on a single user $k^*$ for each OFDM frame, and estimate different users sequentially in time. We achieve this by tuning the reduction matrix $\Um_{k^*} \in \CC^{\Nr\times\Nrfr}$ as will be described later. Then, after standard OFDM processing (see e.g. \cite{sturm2011waveform}) and including noise, the sampled signal when focusing on user $k^*$ is given by
		\begin{align}
			\yv_{k^*}[n, m] &= \Um_{k^*}^\H\bigg(\frac{1}{\sqrt{K}}\sum_{k=1}^{K}h_{k}e^{j2\pi(nT_{\rm 0}\nu_{k} - m\Delta f\tau_{k})}\bv(\phi_{k})\nonumber\\
			&\qquad\qquad\;\;\av^\H(\phi_{k})\fv(\hat{\phi}_{k})\zeta_{k}[n, m] + \wv[n, m]\bigg)\\
			&= \Um_{k^*}^\H\left(\sum_{k=1}^{K}h_{k}g_{t, k}\bv(\phi_{k})\tilde{\zeta}_{k}[n, m] + \wv[n, m]\right) \label{eq:rx_sampled_inter}\\
			&\approx \Um_{k^*}^\H h_{k^*}g_{t, k^*}\bv(\phi_{k^*})\tilde{\zeta}_{k^*}[n, m] + \Um_{k^*}^\H\wv[n, m] \label{eq:rx_sampled},
		\end{align}
		where in \eqref{eq:rx_sampled_inter} we defined $g_{{\rm tx}, k} \eqdef \frac{1}{\sqrt{K}}\av^\H(\phi_{k})\fv(\hat{\phi}_{k})$ and $\tilde{\zeta}_{k}[n, m] \eqdef \zeta_{k}[n, m]e^{j2\pi(n\Tzero\nu_{k} - m\Df\tau_{k})}$, $\wv[n, m] \in \CC^{\Nr}$ is white Gaussian noise with variance $\sigma_{n}^{2}$, and the approximation in \eqref{eq:rx_sampled} follows from designing $\Um_{k^*}$ such that $\|\Um_{k^*}^\H\bv(\phi_{k})\| \approx 0$ for $k \neq k^*$.

	\section{Proposed Tracking Scheme}\label{sec:tracking}
		In this section, we describe the two main components of our tracking scheme. First, we estimate parameters of interest from the sampled signal $\yv_{k^*}[n,m]$. For our tracking scheme, we are interested in the ranges $\{d_{l, k}\}$ and angles of arrival $\{\phi_{l, k}\}$ of the different users, where $l$ indexes measurements sequentially in time. However, due to the nature of our estimator, we also produce estimates of the Doppler frequencies $\{\nu_{l, k}\}$, which could be used for other purposes outside of the scope of this paper. Then, we update the set of predicted angles $\{\phi_{l, k}\}$ based on the history of estimations. The overall procedure is summarized in Algorithm \ref{alg:beam_tracking}.		
		
		\begin{algorithm}[!t]
			\caption{Beam Tracking}
			\begin{algorithmic}
				\INPUT Refresh period $\Delta T$, predicted angles $\{\hat{\phi}_{0, k}\}$ at $t=0$.
				\State $l \gets 0$
				\Loop
					\For{$k^{*}=1,\dots,K$}\footnotemark
					\State Transmit beamformed OFDM frames towards \\
					\hspace{4em} all directions $\{\hat{\phi}_{l, k}\}$ in the predicted set \\
					\hspace{4em} as shown in Section \ref{sec:sys_model}.
					\State Obtain estimate $(\check{d}_{l, k^{*}}, \check{\phi}_{l,k^{*}})$ as shown in \\
					\hspace{4em}Section \ref{ssec:receiver}.
					\State Predict $\hat{\phi}_{l+1, k^{*}}$ as shown in Section \ref{ssec:pred_update}.
					\EndFor
					\State $l \gets l + 1$
					\State Radar receiver idle until $t=l\Delta T$
				\EndLoop
			\end{algorithmic}
			\label{alg:beam_tracking}
		\end{algorithm}
		\footnotetext{Iteration $k^{*}$ of the inner loop corresponds to $t\in[l\Delta T + (k^{*}-1)N\Tzero, l\Delta T + k^{*}N\Tzero)$.}

		\subsection{Radar Parameter Estimation}\label{ssec:receiver}
		For this section, we will focus on estimating the parameters of user $k^*$ at the $l$-th measurement round and will drop the indices to simplify notation. The measurement model is extensively described in \cite{BeamRefinement}, therefore only a brief summary will be provided here. 
		
		We define the reduction matrix $\Um$ as
		\begin{align}
			\Um = \Dm(\hat{\phi})\Psim,
		\end{align} 
		where $\Dm(\hat{\phi}) \eqdef \diag(1, e^{j\pi\sin(\hat{\phi})}, \dots, e^{j\pi(\Nr-1)\sin(\hat{\phi})})$ is a tunable network of phase shifters and $\Psim$ is a fixed network of beamformers pointing towards the broadside direction (i.e. $\phi=0$) , satisfying $\Um^\H\Um = \frac{1}{\Nrfr}\Id_{\Nrfr}$, where $\Id_{N}$ is the identity matrix of rank $N$. In particular, we let the columns of $\Psim$ be obtained as the first $\Nrfr$ Slepian sequences of length $\Nr$ and any user defined time-bandwidth product. A brief description of Slepian sequences can be found in \cite{BeamRefinement} and more extensive treatment in \cite{StoicaSpectralAnalysis}.
		
		Using this design, we can obtain the sample covariance matrix from $N\times M$ samples
		\begin{align}\label{eq:BS_MUSIC}
			\hat{\Rm} &= \frac{1}{NM}\sum_{n=0}^{N-1}\sum_{m=0}^{M-1}\yv[n, m]\yv^\H[n, m]  \nonumber\\
			&\approx P_{\rm tx}\abs{h}^{2}\abs{g_{\rm tx}}^{2}\Psim^\H\bv(\phi')\bv^\H(\phi')\Psim + \sigma_{n}^{2}\Id_{\Nrfr},
		\end{align}
		where $P_{\rm tx} \eqdef \EE\left[\abs{\zeta[n,m]}^2\right]$  is the average power of the transmitted signal and $\phi' \eqdef \sin^{-1}(\sin(\phi) -\sin(\hat{\phi}))$. Given the structure of $\hat{\Rm}$, we can process it via spectral methods such as MUSIC \cite{StoicaSpectralAnalysis} in order to obtain $\check{\phi}$ as a super-resolution estimate of $\phi$.
		
		Finally, we can use the signal model in \eqref{eq:rx_sampled} together with our estimate of $\phi$ to obtain a least squares estimate of $\tau$ and $\nu$. To do so, we solve
		\begin{align}\label{eq:NLS}
			(\check{\tau}, \check{\nu}) = \underset{(\tau, \nu)}{\arg\min}\;\; \sum_{n,m}\left\|\yv[n, m] - hg_{\rm tx}\Um^\H\bv(\check{\phi})\tilde{\zeta}[n, m]\right\|_{2}^{2}\;\;,
		\end{align}
		which, as shown in \cite{BeamRefinement}, is equivalent to solving
		\begin{align}\label{eq:NLS-Last}
			&(\check{\tau}, \check{\nu}) = \underset{(\tau, \nu)}{\arg\max}\;\; \left|\sum_{n,m}y'[n,m]\zeta^{*}[n,m]e^{-j2\pi(n\Tzero\nu - m\Df\tau)}\right|,
		\end{align}
		where $y'[n, m]\eqdef \frac{\bv^\H(\check{\phi})\Um\yv[n, m]}{\bv^\H(\check{\phi})\Um\Um^\H\bv(\check{\phi})}$. Problem \eqref{eq:NLS-Last} can be efficiently solved by applying FFT and finding a peak in a 2D grid. We can estimate the range $d$ as
		\begin{align}
			\check{d} = \frac{c\check{\tau}}{2},
		\end{align}
		where $c$ is the speed of light.
		
		\subsection{Next Angle Prediction}\label{ssec:pred_update}
		In this section, we summarize the key points of the beam prediction step in \cite{LiuLetter3} and propose a combination of their simple tracking equations with the radar receiver introduced in Section \ref{ssec:receiver} that yields an overall improved performance. By focusing still on a single user, we will introduce the time index $l$ to capture the prediction based on the past observations. 
		
		The approach in \cite{LiuLetter3} aims to predict the coordinates of the user of interest given the kinematic state in the last three measurement epochs. In particular, focusing on the $x$ coordinate,
		\begin{align}\label{eq:kinematics}
			\begin{cases}
				&x_{l+1} - x_{l} = v_{x, l}\Delta T + a_{x, l}\Delta T^{2}/2,\\
				&x_{l} - x_{l-1} = v_{x, l-1}\Delta T + a_{x, l-1}\Delta T^{2}/2,\\
				&x_{l-1} - x_{l-2} = v_{x, l-2}\Delta T + a_{x, l-2}\Delta T^{2}/2,\\
				&v_{x, l} - v_{x, l-1}= a_{x, l-1}\Delta T,\\
				&v_{x, l-1} - v_{x, l-2} = a_{x, l-2}\Delta T
			\end{cases}
		\end{align}
		where $x_{l}$, $v_{x, l}$ and $a_{x, l}$ are respectively the $x$-th coordinate, the velocity and the acceleration in the $x$-th coordinate at the $l$-th measurement step, and $\Delta T$ is the interval between measurements corresponding to the same user. In this work, we choose $\Delta T \gg \Tzero$ for two reasons. First, since different users are estimated sequentially in time, the number of users this system can accommodate is upper bounded by $\lfloor\Delta T/\Tzero\rfloor$, where $\lfloor x \rfloor$ indicates the largest integer not greater than $x$. Second, $\Tzero$ can be too short for any non-negligible motion to happen. Therefore, using a much longer $\Delta T$ does not significantly affect performance, while greatly simplifying hardware operation by reducing the rate at which beams must be reconfigured. 
		
		By setting the assumption that the acceleration does not significantly change in three measurement steps (i.e. $a_{x, l-2} \approx a_{x, l-1} \approx a_{x, l}$), the system of equations in \eqref{eq:kinematics} can be solved for $x_{l+1}$ yielding
		\begin{align}
			\hat{x}_{l+1} = 3x_{l} - 3x_{l-1} + x_{l-2} \approx 3\check{x}_{l} - 3\check{x}_{l-1} + \check{x}_{l-2},
		\end{align}
		where, keeping our notation consistent, $\check{x}$ represents an estimated value (e.g. by following the processing in Section \ref{ssec:receiver})\footnote{Notice that estimates of $x$ and $y$ can be directly obtained from estimates of $d$ and $\phi$ using basic trigonometry.}, and $\hat{x}$ represents a predicted value.
		
		The derivation holds verbatim for the $y$-th coordinate, so we can obtain a prediction for the next AoA from the relation between the $x$ and $y$ coordinates as
		\begin{align}\label{eq:Liu-Prediction}
			\hat{\phi}_{l+1} = \tan^{-1}\left(\frac{3\check{x}_{l} - 3\check{x}_{l-1} + \check{x}_{l-2}}{3\check{y}_{l}-3\check{y}_{l-1}+\check{y}_{l-2}}\right).
		\end{align}

	\section{Numerical Results}
		We compare our approach with the one proposed in \cite{LiuLetter3} that assumes a fully digital receiver (i.e. $\Nrfr=\Nr$) and considers estimation via local linear approximation \cite[Section III.B]{LiuLetter3}.
		We remark here that our method, by operating in the beam-space domain, overcomes two major issues associated with fully digital architectures. First, it greatly reduces hardware complexity since the wideband signal is sampled by only $\Nrf$ A/D converters, where we can in general have $\Nrf \ll \Na$ as will be shown in this section. Second, it should be noted that the raw samples at the antenna outputs are extremely noisy due to the large pathloss of the two-way \ac{mmWave} channel. Our method projects the high dimensional signal into a suitable low dimensional subspace where most of the information content is preserved but noise from uninteresting directions is rejected, thus resulting in a much higher SNR per sample. Slepian sequences provide a systematic way to define such a subspace by procuring an orthonormal basis upon which signals coming from directions of interest have maximal projection, whereas those coming from any other angular region are approximately orthogonal.
		We also note that we had to assume perfect matched filtering for the method in \cite{LiuLetter3} to perform adequately, whereas our method is robust to small processing errors (e.g. those caused by discrete sampling in time and frequency). A detailed description of what this assumption entails can be found in Appendix \ref{app:PerfectMF}. For the sake of completeness, we also include performance evaluation for the method in \cite{LiuLetter3} when no perfect matched filtering is assumed. Furthermore, for ease of evaluation, in the presented results we focus on a single user scenario (i.e. $K=1$). Table \ref{tab:System-Parameters} summarizes the parameters used in our simulations. 
		
		\renewcommand{\arraystretch}{1.3}
		\begin{table}
			\caption{System parameters}
			\centering
			\begin{tabular}{|c|c|}
				\hline
				$N=64$ & $M=512$ \\ \hline
				$\Ntx=\Nr=64$ & $\Nrft=\Nrfr=4$ \\ \hline
				$f_{\rm c}=$ \SI{60}{\giga\hertz} & $\Df=$ \SI{1}{\mega\hertz} \\ \hline
				$d_{\rm max} = $ \SI{100}{\meter} & $v_{\rm max} =$ \SI{30}{\meter/\second}\\ \hline
				$P_{\rm TX}$ = \SI{50}{\milli\watt} & $\sigma_{\rm rcs}=$ \SI{20}{\deci\belsquaremeter} \\ \hline
				Noise PSD: $N_{0}$ = \SI{2e-21}{\watt/\hertz} & $\Delta T= $ \SI{100}{\milli\second} \\ \hline
				
			\end{tabular}
			\label{tab:System-Parameters}
		\end{table}
		
		\begin{figure}
			\centering
			\includegraphics[width=.9\linewidth]{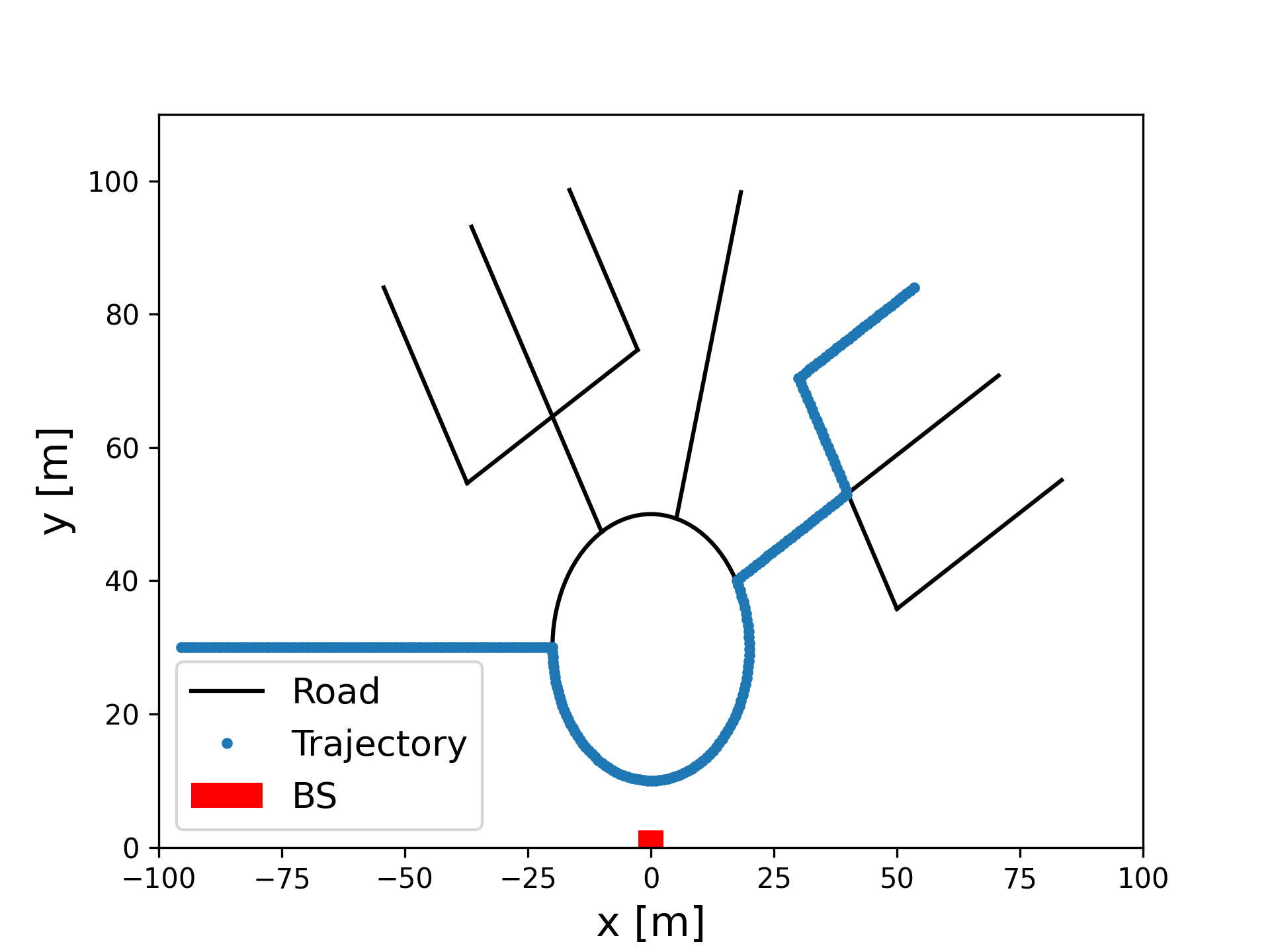}
			\caption{Main road scenario and trajectory considered.}
			\label{fig:trajectory}
		\end{figure}
	
		\begin{figure}
		\begin{center}
			\resizebox{1.\linewidth}{!}{\input{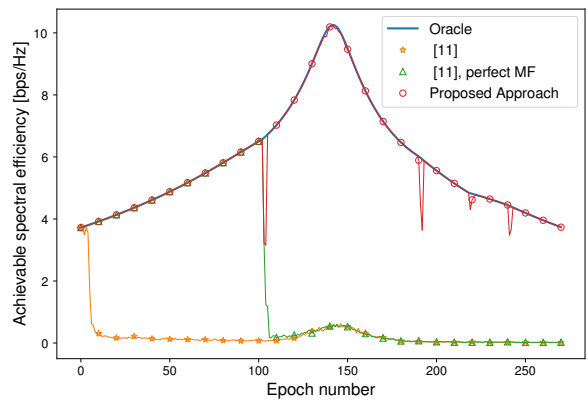}}
		\end{center}
				\caption{Achievable spectral efficiency as a function of the time step for the selected trajectory.}
				\label{fig:SE_Single_Traj}
		\end{figure}
	
		\begin{figure}
		\begin{center}
			\resizebox{1\linewidth}{!}{\input{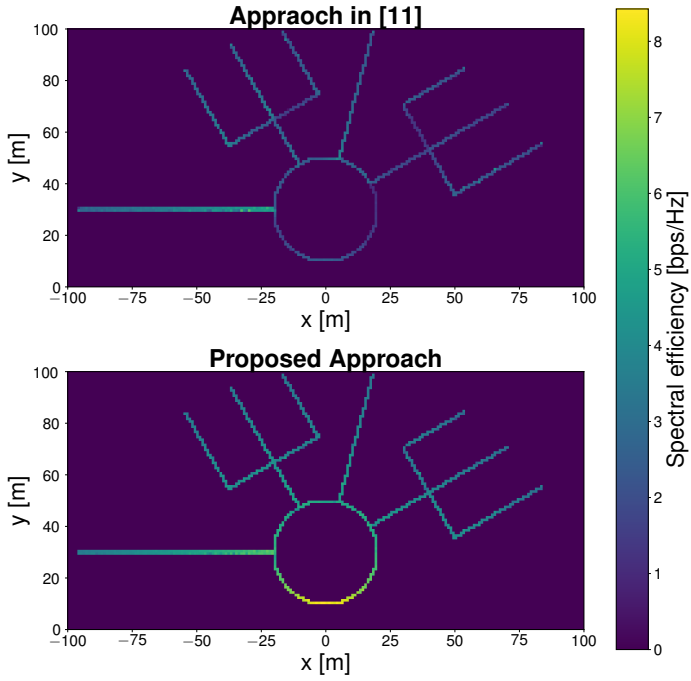}}
		\end{center}
			\caption{Achievable spectral efficiency as a function of the position in the road for each of the two trackers considered.}
			\label{fig:Heatmaps}
		\end{figure}

		To verify our results in arbitrary road geometries, we consider the road structure shown in \figurename\ref{fig:trajectory}. In particular, for our first result, we fix the path shown in blue and study the performance of the two trackers in terms of achievable spectral efficiency as a function of the time step, averaged over 500 Monte Carlo simulations. The achievable spectral efficiency assumes a single antenna user, and is computed as 
		\begin{align}\label{eq:AverageSE}
			\eta_{l} \approx \EE\left[\log_{2}\left(1 + \left( \frac{\lambda}{4\pi d_{l}}\right)^{2}\frac{P_{\rm tx}\vert g_{{\rm tx}, l}\vert^{2}}{N_{0}(M\Df)}\right)\right],
		\end{align}
		where the approximation follows from using an empirical estimation of the expectation and $N_{0}$ is the noise power spectral density at the user. The results are presented in \figurename\ref{fig:SE_Single_Traj}, and show how our method is able to quickly recover from the non-smooth changes of direction while the approach in \cite{LiuLetter3} loses track of the user and is not able to find it again. 
		
		In order to average results over different mobility patterns, we generate trajectories where a user moves within the road structure defined in \figurename\ref{fig:trajectory}, but now is able to take any of the paths, and where the speed at each time step is sampled from a random process. Since now there is no association between time step and kinematic state, we illustrate performance as a function of the position by means of a heat map. This is shown in \figurename\ref{fig:Heatmaps}, where our method clearly outperforms the baseline in \cite{LiuLetter3} when averaging over an ensemble of different trajectories. In particular, it can be observed that the approach in \cite{LiuLetter3} fails most of the time to track the user in the locations closest to the \ac{BS}, where the angles change at a faster rate, even though the path loss there is small. On the other hand, the performance of our proposed method depends only on the distance (or equivalently the path loss) to the user, but is robust to fast changes in the trajectory.
		
		\begin{figure}
			\centering
			\includegraphics[width=.9\linewidth]{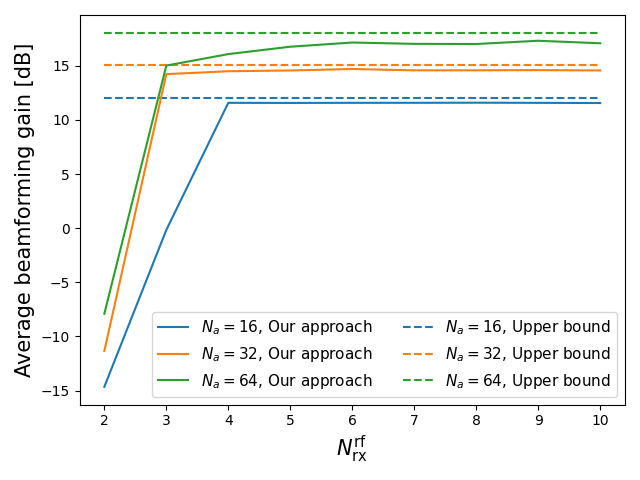}
			\caption{Average beamforming gain as a function of the number of RF chains for different numbers of antenna elements.}
			\label{fig:Perf_per_NRF}
		\end{figure}
		
		Finally, we study the effect of the number of RF chains in the tracking performance. To do so, we generate multiple trajectories and average the beamforming gain obtained at each point of each path for different antenna front end configurations. This is presented in \figurename\ref{fig:Perf_per_NRF}, where it can be seen that as few as 4 RF chains are enough to achieve close to optimal performance. This result suggests that projecting the high dimensional signal at the antenna plane onto a much lower dimensional space preserves most of the information necessary for estimation, provided that the considered subspace is adequately selected. The figure also shows how the gap between the achieved results and the upper bound becomes larger when the number of antennas increases. This can be explained by the fact that larger arrays generate narrower beams, resulting on a higher sensitivity to small pointing errors. 
				
	\section{Conclusions}
		In this paper, we have shown an efficient method to perform beam tracking in OFDM \ac{ISAC} systems, based on the combination of an advanced \ac{HDA} receiver and a simple tracking equation. Our close to optimal results indicate that large savings in computation and hardware complexity can be obtained without sacrificing performance. Moreover, the validation of our results in  complicated road geometries suggests that our method is general enough to perform well in realistic mobility scenarios. 
	
	\bibliographystyle{IEEEtran}
	\bibliography{estcomm,IEEEabrv,bibliography}
	
	\begin{appendices}
		\section{The perfect matched filtering assumption}\label{app:PerfectMF}
		In this appendix, we illustrate the limitation of the assumption on perfect matched filtering frequently used in recent works. To that aim, let us focus for simplicity in a single path channel model, such that the time-varying channel matrix is given by
		\begin{align}
			\Hm (t, \tau) =  h_{0} \bv(\phi_{0}) \av^\H(\phi_{0})\delta(\tau-\tau_{0})  e^{j2\pi \nu_{0} t},
		\end{align}
		where $h_{0}$, $\phi_{0}$, $\tau_{0}$ and $\nu_{0}$ represent respectively the channel coefficient, AoA, delay and Doppler associated with a single user. Let us consider 
		a generic transmitted signal $\sv(t)\eqdef \fv s(t)$ time limited to the interval $t\in[0, T]$, where $\fv$ is a beamforming vector, such that the received signal in the absence of noise is
		\begin{align}
			\rv(t) = h_{0}g(\phi_{0})\bv(\phi_{0})s(t-\tau_{0})e^{j2\pi\nu_{0}t} ,
		\end{align}
		where $g(\phi_{0}) \eqdef \av^\H(\phi_{0})\fv$. The (ideal) matched filtering process would then consist in processing the received signal as
		\begin{align}
			\rv &= \int_{-\infty}^{\infty}\rv(t')s^*(t'-(t-T))e^{-j2\pi\nu t'}dt'\bigg|_{\substack{t=T + \tau_{0} \\ \nu=\nu_{0}}}\\
			&=h_{0}g(\phi_{0})\bv(\phi_{0})\nonumber\\
			&\qquad\int_{-\infty}^{\infty}s(t'-\tau_{0})s^*(t'-(t-T))e^{-j2\pi(\nu_{0} - \nu) t'}dt'\bigg|_{\substack{t=T + \tau_{0} \\ \nu=\nu_{0}}}\\
			&=h_{0}g(\phi_{0})\bv(\phi_{0})\int_{-\infty}^{\infty}s(t'-\tau_{0})s^*(t'-\tau_{0})dt'\\
			&=h_{0}\Ec_{s}g(\phi_{0})\bv(\phi_{0}), \label{eq:perfectMF}
		\end{align}
		where $\Ec_{s}\eqdef\int_{-\infty}^{\infty}\vert s(t) \vert^{2}dt$ is the energy of signal $s(t)$. Recent works consider a noisy version of \eqref{eq:perfectMF} as a model for the sampled signal. However, as illustrated here, explicit knowledge of the delay and Doppler of the channel was required to arrive to this expression. This assumption becomes unrealistic in mobility settings where delay and Doppler might be time varying magnitudes. In practice, most systems sample a grid of $(\tau, \nu)$ tuples and find the pair maximizes the norm of the output of the matched filter. Let us refer to such a pair as $(\hat{\tau}, \hat{\nu})$, where $(\hat{\tau}, \hat{\nu}) \neq (\tau_{0}, \nu_{0})$ almost surely due to the finiteness of the grid. Then, the exact matched filter output would be given by
		\begin{align}\label{eq:imperfectMF}
			\rv = h_{0}g(\phi_{0})\bv(\phi_{0})\int_{-\infty}^{\infty}s(t'-\tau_{0})s^*(t'-\hat{\tau})e^{j2\pi(\nu_{0}-\hat{\nu})t'}dt',
		\end{align}
		which shows that the term $\Ec_{s}$ should be replaced with an expression depending on $\Delta \tau \eqdef \tau_{0} - \hat{\tau}$ and $\Delta\nu \eqdef \nu_{0}-\hat{\nu}$. Notice also that $\Delta\tau$ and $\Delta\nu$ can vary quickly in time since $\tau_{0}$ ($\nu_{0}$) changes smoothly and $\hat{\tau}$ ($\hat{\nu}$) changes in discrete steps. When using model \eqref{eq:imperfectMF} in lieu of \eqref{eq:perfectMF}, many of the works in the literature become unusable. In particular, works that consider that the sampled signal is described by a statistical distribution that is a perfectly known function of the AoA and the channel coefficient fail to consider this difficult to model phenomenon. Also, methods that take derivatives of the model with respect to the parameters in order to linearize the functional dependence, like the baseline used in this paper, have typically considered the formulation in \eqref{eq:perfectMF}, resulting in simple expressions that are however invalid once the perfect matched filtering assumption is lifted.
	\end{appendices}

\end{document}